\documentclass[11pt,twocolumn]{article}

\usepackage[utf8]{inputenc}
\usepackage[T1]{fontenc}
\usepackage{amsmath,amssymb,amsthm}
\usepackage{graphicx}
\usepackage{booktabs}
\usepackage{algorithm}
\usepackage{algorithmic}
\usepackage{subcaption}
\usepackage[margin=0.85in]{geometry}
\usepackage{adjustbox}
\usepackage{multirow}
\usepackage{xcolor}
\usepackage{float}
\usepackage{enumitem}
\usepackage{natbib}
\usepackage{url}
\usepackage[breaklinks,colorlinks,citecolor=blue!70!black,linkcolor=blue!60!black,urlcolor=blue!70!black]{hyperref}

\newtheorem{definition}{Definition}
\newtheorem{proposition}{Proposition}
\newtheorem{remark}{Remark}

\title{\textbf{PROXIMA: Proxy Metric Validation with Segment-Level\\Fragility Detection for Online Controlled Experiments}}

\author{
    Avinash Amudala \\
    Rochester Institute of Technology \\
    \texttt{aa9429@g.rit.edu}
}

\date{April 2026}

\begin{document}

\maketitle

\begin{abstract}
Online A/B testing at scale relies on proxy metrics---short-term, easily-measured signals used in place of slow-moving long-term outcomes.
When the proxy--outcome relationship is heterogeneous across user segments, aggregate correlation can mask directional failures akin to Simpson's Paradox, leading to costly ship/no-ship errors.
We introduce \textbf{PROXIMA} (\textbf{Prox}y Metr\textbf{i}c Validation Fra\textbf{m}ework for Online Experiments), a lightweight diagnostic framework that scores proxy reliability through a composite of three complementary dimensions: normalised effect correlation, directional accuracy, and segment-level fragility rate.
Unlike surrogate-index approaches that predict long-term treatment effects, PROXIMA directly audits whether a candidate proxy leads to correct launch decisions and flags the user segments where it fails.
We validate PROXIMA on two public datasets---the Criteo Uplift corpus (14M observations, advertising) and KuaiRec (7K users, video recommendation)---using 80 simulated A/B tests.
Early engagement metrics achieve a composite reliability of 0.80 on Criteo and 0.62 on KuaiRec, yielding 98.4\% average decision agreement with an oracle policy.
Fragility analysis reveals that recommendation domains exhibit substantially higher segment-level heterogeneity (68\% fragility) than advertising (13\%), yet directional accuracy remains above 96\% in both cases.
A sensitivity analysis over the weight space confirms that no single component suffices and that the composite provides substantially better discrimination between reliable and unreliable proxies than correlation alone.
Code and reproduction scripts are available at \url{https://github.com/Avinash-Amudala/PROXIMA}.
\end{abstract}


\section{Introduction}
\label{sec:intro}

Online controlled experiments (OCEs)---commonly called A/B tests---are the gold standard for data-driven product decisions at technology companies \citep{kohavi2009controlled,kohavi2013online,larsen2024statistical}.
A typical experiment randomly assigns users to treatment and control variants, then measures the difference in an \emph{Overall Evaluation Criterion} (OEC) such as long-term retention or lifetime revenue.

\paragraph{The proxy metric dilemma.}
Measuring the true long-term OEC requires weeks or months of observation, which is incompatible with the pace of iterative product development.
In practice, organisations substitute a \emph{proxy metric}---an early, fast-moving signal such as click-through rate, session count, or short-horizon engagement---that is believed to correlate with the OEC \citep{deng2017statistical,hagar2023choosing}.
When the correlation holds, proxies allow decisions in days rather than months.

\paragraph{Failure modes.}
Proxy metrics can fail silently.
A proxy that is positively correlated with the long-term outcome \emph{in aggregate} may give the opposite signal in specific user segments---a manifestation of Simpson's Paradox \citep{simpson1951interpretation,pearl2014understanding}.
For instance, an increase in video watch time may predict retention for new users but signal content fatigue (and eventual churn) for power users.
Relying on the aggregate proxy then leads to shipping changes that harm a substantial sub-population.
Existing surrogate-validation methods \citep{athey2019surrogate,zhang2023evaluating} focus on predicting long-term effects from short-term data but do not explicitly audit segment-level fragility.

\paragraph{Contributions.}
We present \textbf{PROXIMA}, a framework that goes beyond correlation to score proxy reliability along three dimensions:

\begin{enumerate}[leftmargin=*,topsep=2pt,itemsep=1pt]
    \item \textbf{Composite Reliability Score.} A normalised, interpretable metric that combines effect correlation, directional accuracy, and segment-level fragility rate, with all components mapped to $[0,1]$ (Section~\ref{sec:composite}).
    \item \textbf{Segment-Level Fragility Detection.} An automated procedure that detects segments where the proxy effect sign contradicts the global long-term direction---operationalising Simpson's Paradox detection for A/B testing (Section~\ref{sec:fragility}).
    \item \textbf{Decision Simulation Framework.} A counterfactual analysis comparing proxy-based ship/no-ship decisions to an oracle with perfect long-term knowledge, yielding win rate, false-positive/negative rates, and decision regret (Section~\ref{sec:decision_sim}).
    \item \textbf{Multi-Domain Validation.} Evaluation on 14M+ observations spanning advertising (Criteo) and recommendation (KuaiRec) domains, showing 98.4\% oracle agreement and revealing cross-domain differences in fragility (Section~\ref{sec:results}).
\end{enumerate}

\section{Related Work}
\label{sec:related}

\subsection{Online Controlled Experiments at Scale}

The modern practice of web-scale A/B testing was established by the seminal surveys of \citet{kohavi2009controlled} and \citet{kohavi2013online}, and has since been adopted at virtually every major technology company.
\citet{larsen2024statistical} provide a comprehensive review of the statistical challenges that arise in OCEs---including interference, metric selection, multiple testing, and the proxy-metric problem---and call for deeper academic--industry collaboration.
PROXIMA addresses the \emph{metric selection} challenge by providing a principled diagnostic for proxy reliability.

\subsection{Surrogate Metrics and the Surrogate Index}

The statistical literature on surrogate endpoints dates to the Prentice criteria \citep{prentice1989surrogate}, which formalise conditions under which a treatment effect on a surrogate can substitute for the effect on the true endpoint.
\citet{athey2019surrogate} introduced the \emph{surrogate index} for causal inference: a weighted combination of short-term outcomes that, under a surrogacy assumption, recovers the average treatment effect on the long-term outcome.
\citet{zhang2023evaluating} evaluated the surrogate index at Netflix using 1{,}098 test arms from 200 A/B tests, finding 95\% consistency between 14-day surrogate predictions and 63-day outcomes, with 79\% precision for launch decisions.

These methods aim to \emph{predict} long-term effects from short-term data.
PROXIMA complements this line of work by \emph{auditing} an existing proxy: rather than constructing an optimal surrogate, it scores how reliable a given proxy is and detects the segments where it fails.

\subsection{Optimal Proxy Metric Construction}

\citet{hagar2023choosing} frame proxy selection as a portfolio optimisation problem, showing that the optimal proxy depends on the experiment's sample size and noise level.
\citet{liu2023pareto} extend this to multi-objective optimisation, identifying Pareto-optimal proxies that jointly maximise prediction accuracy and short-term sensitivity; their constructed proxies were eight times more sensitive than the north-star metric.
\citet{hagar2023cuped} revisit CUPED \citep{deng2013improving} as a variance reduction framework and extend it to in-experiment data.

PROXIMA is agnostic to how the proxy was chosen: it can evaluate any candidate, whether hand-picked by a domain expert, selected by a portfolio optimiser, or constructed via the surrogate index.

\subsection{Simpson's Paradox and Subgroup Analysis}

Simpson's Paradox---the reversal of an association when data is stratified---has a rich statistical history \citep{simpson1951interpretation,pearl2014understanding,blyth1972simpson}.
In the context of OCEs, Statsig's \emph{Differential Impact Detection} \citep{statsig2024did} automatically flags experiments with heterogeneous sub-population impacts using segment-wise Welch's $t$-tests with Bonferroni correction.
\citet{teng2026deparadox} recently proposed the De-paradox Tree, a kernel-based recursive-partitioning algorithm that uncovers nested paradoxical associations.
Netflix's \texttt{sherlock} library \citep{netflix2023sherlock} uses doubly-robust causal machine learning to discover segments that would benefit from or be harmed by a treatment.

PROXIMA differs from these approaches in scope: rather than testing for heterogeneous treatment effects on a \emph{single} outcome, it tests whether the \emph{proxy--outcome directional relationship} is heterogeneous across segments.

\subsection{Treatment Effect Heterogeneity}

Causal forests \citep{wager2018estimation} and meta-learners \citep{kunzel2019metalearners,athey2016recursive} estimate Conditional Average Treatment Effects (CATEs) at the individual or segment level.
These methods are powerful for personalisation but address a different question: ``Who benefits most?'' versus PROXIMA's ``In which segments does the proxy give the wrong signal?''

\subsection{Metric Integrity and Goodhart's Law}

\citet{goodhart1984problems} observed that measures cease to be reliable once they become targets.
In online experimentation, this manifests as metric gaming---optimising click-through rate with clickbait, for example, while degrading satisfaction \citep{manzi2012uncontrolled}.
PROXIMA's fragility analysis can serve as an early-warning system: a proxy with high aggregate correlation but high segment-level fragility may indicate Goodhart-style degradation in specific user populations.

\subsection{Sequential Testing and Early Stopping}

Group sequential designs \citep{jennison1999group} and always-valid confidence sequences \citep{howard2021time} allow experimenters to monitor a metric continuously and stop early.
\citet{johari2017peeking} address the peeking problem in A/B tests.
These methods assume the monitored metric is the true outcome; PROXIMA addresses the orthogonal question of whether a \emph{different, faster} metric can substitute for the slow outcome.

\section{Method}
\label{sec:method}

\subsection{System Overview}

\begin{figure}[t]
\centering
\includegraphics[width=\columnwidth]{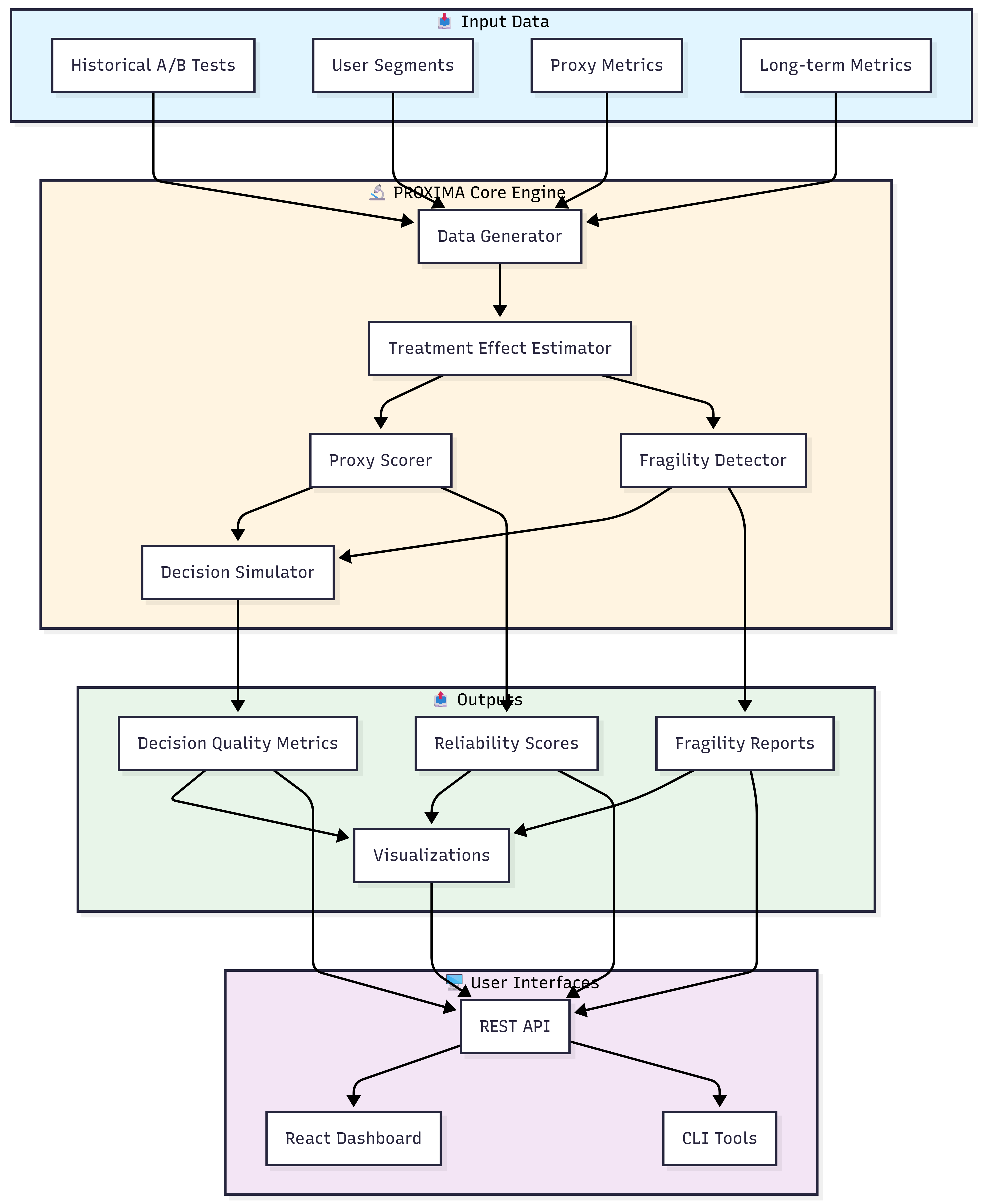}
\caption{High-level architecture of PROXIMA.  Historical experiment data is processed by the core engine (composite scoring, fragility detection, decision simulation) and surfaced via an API and dashboard.}
\label{fig:architecture}
\end{figure}

Figure~\ref{fig:architecture} provides a high-level view of the PROXIMA pipeline.
Historical A/B test data flows through the core scoring engine, which computes composite reliability, detects fragile segments, and simulates decisions.
Results are exposed via an API and interactive dashboard.

\subsection{Problem Setup}
\label{sec:setup}

Consider a corpus of $E$ historical A/B tests.
In experiment $e$ ($e = 1, \ldots, E$), each user $i$ is randomly assigned to treatment ($T_i = 1$) or control ($T_i = 0$) and belongs to a pre-defined segment $S_i \in \mathcal{S}$.
We observe an early proxy metric $Y_i^{\text{proxy}}$ and a long-term outcome $Y_i^{\text{long}}$.
The experiment-level average treatment effects are:
\begin{equation}
    \tau_m^{e} \;=\; \bar{Y}_m^{e,\,\text{treat}} - \bar{Y}_m^{e,\,\text{ctrl}}, \quad m \in \{\text{proxy}, \text{long}\}
    \label{eq:ate}
\end{equation}
and the segment-level effects are $\tau_m^{e,s}$ for segment $s$.

\begin{definition}[Proxy Reliability]
A proxy metric is \textbf{reliable} if (i) its treatment effects correlate with long-term effects across experiments, (ii) it leads to the same ship/no-ship decision as the long-term metric in most experiments, and (iii) these properties hold uniformly across user segments.
\end{definition}

\noindent PROXIMA formalises this intuition through a composite score that penalises failures on any of the three dimensions.

\subsection{Composite Reliability Score}
\label{sec:composite}

Let $\boldsymbol{\tau}^{\text{proxy}} = (\tau_{\text{proxy}}^1, \ldots, \tau_{\text{proxy}}^E)$ and $\boldsymbol{\tau}^{\text{long}} = (\tau_{\text{long}}^1, \ldots, \tau_{\text{long}}^E)$ be the vectors of experiment-level effects.
We define three component scores:

\paragraph{Normalised Effect Correlation.}
\begin{equation}
    C \;=\; \frac{\rho(\boldsymbol{\tau}^{\text{proxy}}, \boldsymbol{\tau}^{\text{long}}) + 1}{2} \;\in\; [0, 1]
    \label{eq:corr_norm}
\end{equation}
where $\rho$ denotes the Pearson correlation.
Mapping from $[-1,1]$ to $[0,1]$ ensures commensurability with the other components and prevents negative reliability scores.

\paragraph{Directional Accuracy.}
\begin{equation}
    \text{DA} \;=\; \frac{1}{E} \sum_{e=1}^{E} \mathbb{I}\!\left[\operatorname{sign}(\tau_{\text{proxy}}^e) = \operatorname{sign}(\tau_{\text{long}}^e)\right] \;\in\; [0, 1]
    \label{eq:da}
\end{equation}
This captures the decision-making utility of the proxy: a practitioner who ships whenever the proxy effect is positive makes the same decision as an oracle whenever the signs agree.

\paragraph{Fragility Rate.}
\begin{equation}
    \text{FR} \;=\; \frac{1}{\sum_e |\mathcal{S}_e|} \sum_{e=1}^{E} \sum_{s \in \mathcal{S}_e} \mathbb{I}\!\left[\operatorname{sign}(\tau_{\text{proxy}}^{e,s}) \neq \operatorname{sign}(\tau_{\text{long}}^{e})\right]
    \label{eq:fr}
\end{equation}
where $\mathcal{S}_e$ is the set of segments in experiment $e$.
A segment is \emph{fragile} if the proxy's segment-level effect has the opposite sign of the global long-term effect.
Fragility captures Simpson's Paradox--like heterogeneity: the proxy looks correct globally but misleads in specific segments.

\paragraph{Composite Score.}
The reliability score is a weighted combination:
\begin{equation}
    R \;=\; w_C \cdot C \;+\; w_{\text{DA}} \cdot \text{DA} \;+\; w_{\text{FR}} \cdot (1 - \text{FR})
    \label{eq:composite}
\end{equation}
with $w_C + w_{\text{DA}} + w_{\text{FR}} = 1$ and all weights non-negative.
Since each component lies in $[0,1]$, the composite score $R \in [0,1]$.

\begin{proposition}[Properties of $R$]
\label{prop:properties}
The composite reliability score $R$ satisfies:
\begin{enumerate}[leftmargin=*,topsep=2pt,itemsep=1pt]
    \item \textbf{Boundedness:} $R \in [0, 1]$.
    \item \textbf{Monotonicity:} $R$ is non-decreasing in $C$, non-decreasing in DA, and non-increasing in FR.
    \item \textbf{Interpretability:} $R = 1$ iff perfect correlation ($\rho = 1$), perfect directional accuracy, and zero fragility.
    \item \textbf{Degradation:} $R = 0$ iff anti-correlated ($\rho = -1$), zero directional accuracy, and total fragility.
\end{enumerate}
\end{proposition}

\paragraph{Weight Selection.}
We set $w_C = 0.6$, $w_{\text{DA}} = 0.2$, $w_{\text{FR}} = 0.2$.
The rationale is that effect correlation is the primary signal of proxy quality, while directional accuracy and fragility rate act as regularisers that guard against two distinct failure modes (wrong global direction and wrong segment-level direction).
We study sensitivity to these weights in Section~\ref{sec:sensitivity}.

\subsection{Fragility Detection}
\label{sec:fragility}

For a given proxy metric and a segmentation $\mathcal{S}$, the fragility detection procedure identifies the specific segments where the proxy is unreliable.

\begin{definition}[Fragile Segment]
Segment $s$ in experiment $e$ is \textbf{fragile} with respect to proxy $p$ if
\[
    \operatorname{sign}(\tau_{p}^{e,s}) \neq \operatorname{sign}(\tau_{\text{long}}^{e})
\]
i.e., the proxy's local treatment effect in that segment points in the opposite direction from the global long-term effect.
\end{definition}

\noindent We aggregate fragility by segment identity across experiments to produce a \emph{fragility profile}: for each segment $s$, we compute the fraction of experiments in which $s$ is fragile.
Segments with consistently high fragility rates are flagged as risk zones.

\begin{remark}
We compare the segment-level proxy effect to the \emph{global} long-term effect rather than the segment-level long-term effect.
This design choice reflects the practical decision setting: a practitioner observes the segment-level proxy signal and asks whether it supports the same ship decision implied by the global long-term outcome.
\end{remark}

\subsection{Statistical Testing}
\label{sec:testing}

For each experiment $e$ and metric $m$, we assess significance via Welch's $t$-test:
\begin{equation}
    t = \frac{\tau_m^e}{\sqrt{s_T^2/n_T + s_C^2/n_C}}
    \label{eq:welch}
\end{equation}
and report effect size via Cohen's $d$:
\begin{equation}
    d = \frac{\tau_m^e}{\sqrt{\frac{(n_T-1)s_T^2 + (n_C-1)s_C^2}{n_T+n_C-2}}}
    \label{eq:cohend}
\end{equation}

Confidence intervals for the composite reliability score are obtained via the nonparametric bootstrap (1{,}000 resamples over experiments), recomputing $R$ from each bootstrap sample of effect pairs.

\subsection{Decision Simulation}
\label{sec:decision_sim}

To quantify the practical impact of proxy choice, we simulate ship/no-ship decisions:

\begin{itemize}[leftmargin=*,topsep=2pt,itemsep=1pt]
    \item \textbf{Proxy policy:} Ship experiment $e$ if $\tau_{\text{proxy}}^e > 0$.
    \item \textbf{Oracle policy:} Ship experiment $e$ if $\tau_{\text{long}}^e > 0$.
\end{itemize}

\noindent We then compute:
\begin{itemize}[leftmargin=*,topsep=2pt,itemsep=1pt]
    \item \textbf{Win rate}: fraction of experiments where proxy and oracle agree.
    \item \textbf{False positive rate (FPR)}: fraction of proxy-shipped experiments that the oracle would not ship.
    \item \textbf{False negative rate (FNR)}: fraction of oracle-shipped experiments that the proxy would not ship.
    \item \textbf{Regret}: mean long-term ATE gap between oracle and proxy decision outcomes.
\end{itemize}

\subsection{End-to-End Workflows}

\begin{figure}[t]
\centering
\begin{subfigure}{0.48\linewidth}
\centering
\includegraphics[width=\linewidth]{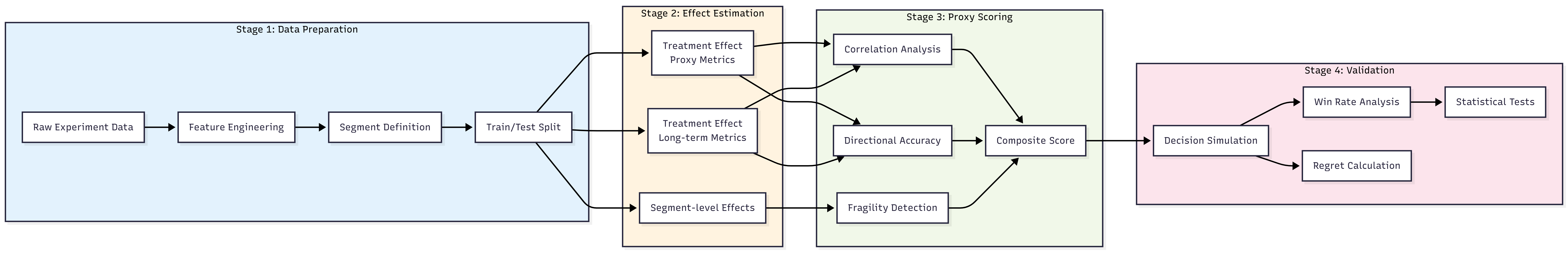}
\caption{Data flow from raw logs to experiment-level proxy and long-term metrics.}
\label{fig:dataflow}
\end{subfigure}
\hfill
\begin{subfigure}{0.48\linewidth}
\centering
\includegraphics[width=\linewidth]{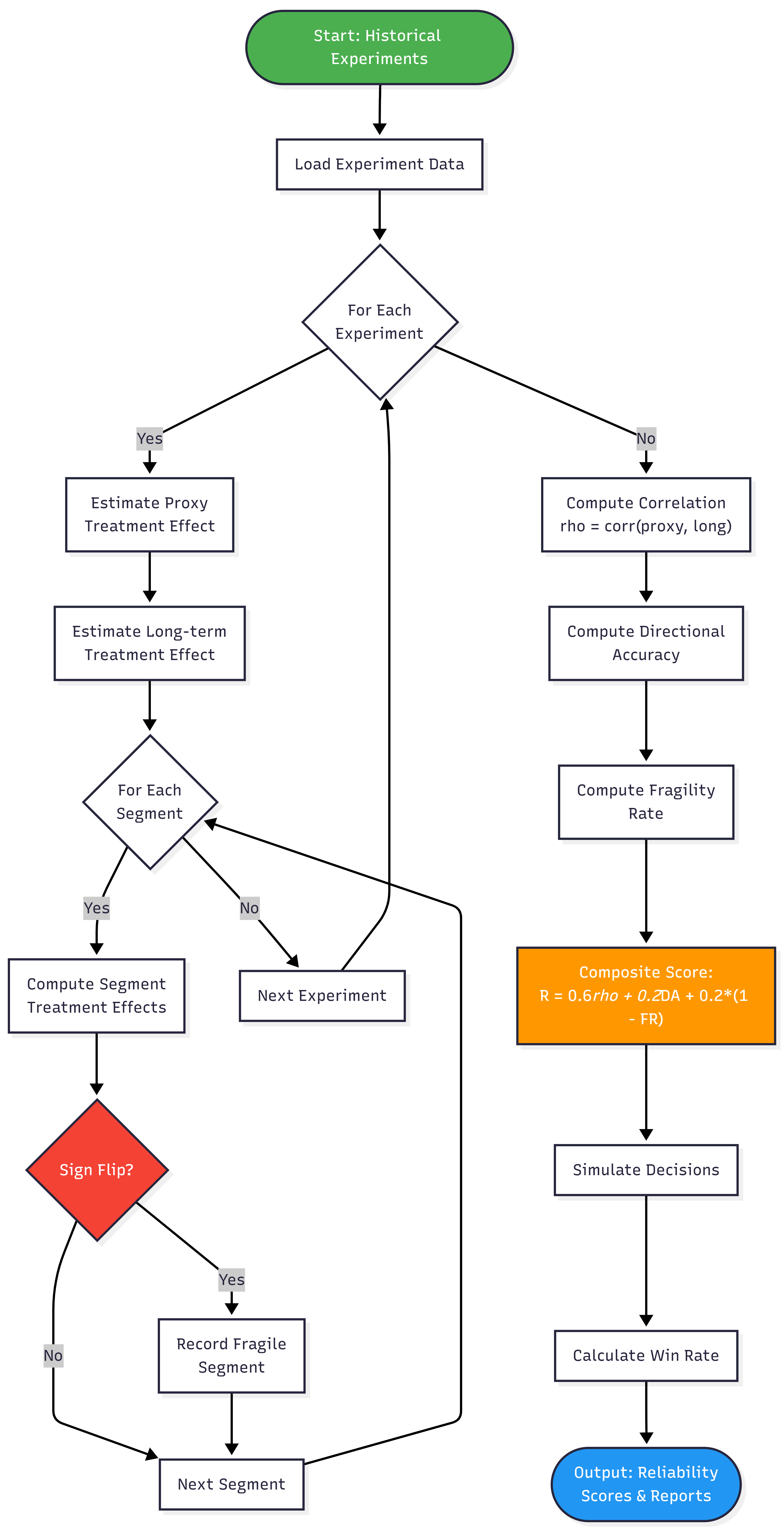}
\caption{Algorithmic flow for composite reliability scoring and fragility detection.}
\label{fig:algo_flow}
\end{subfigure}

\begin{subfigure}{0.48\linewidth}
\centering
\includegraphics[width=\linewidth]{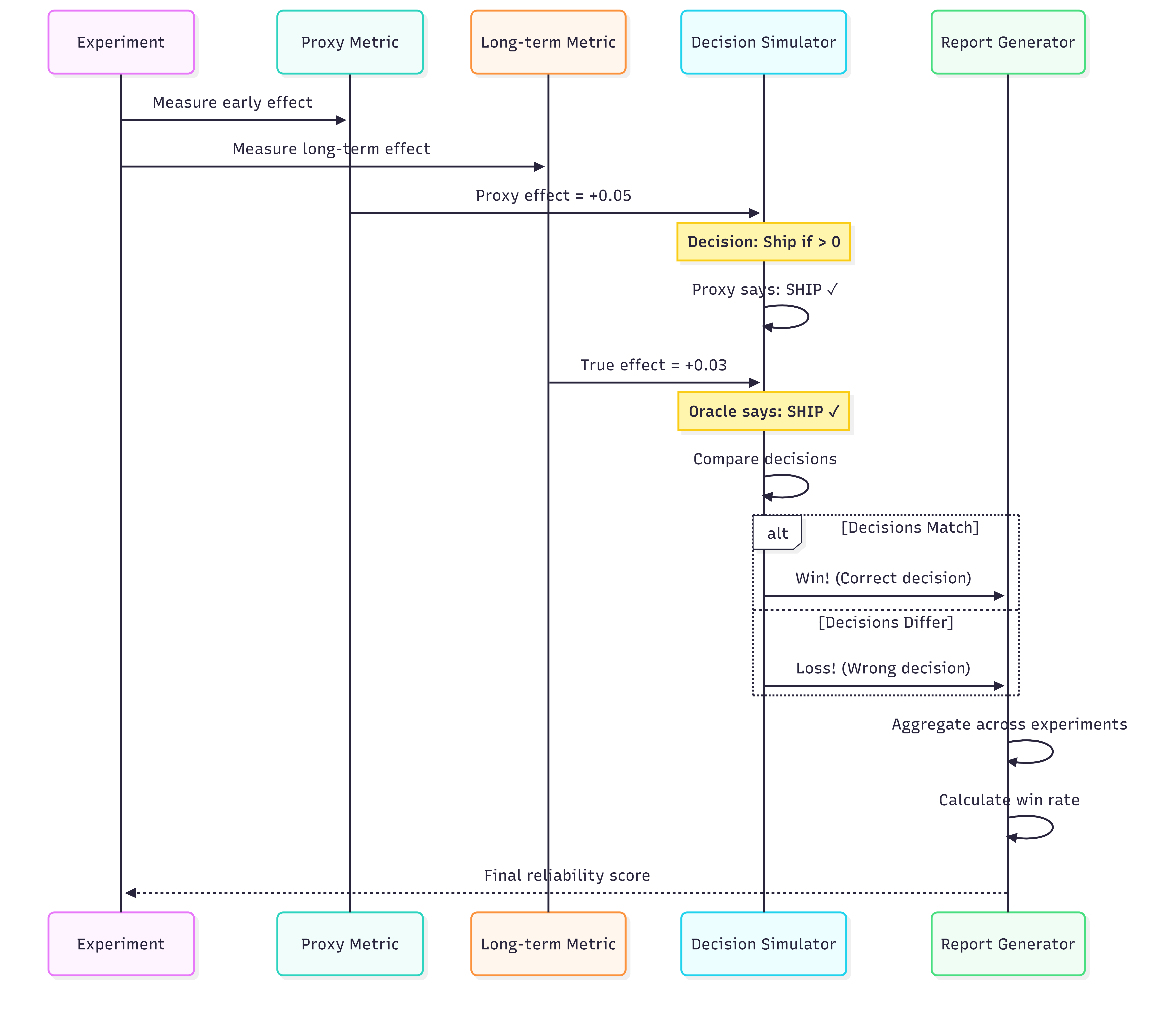}
\caption{Decision simulation pipeline comparing proxy-based and oracle policies.}
\label{fig:simulation_flow}
\end{subfigure}
\hfill
\begin{subfigure}{0.48\linewidth}
\centering
\includegraphics[width=\linewidth]{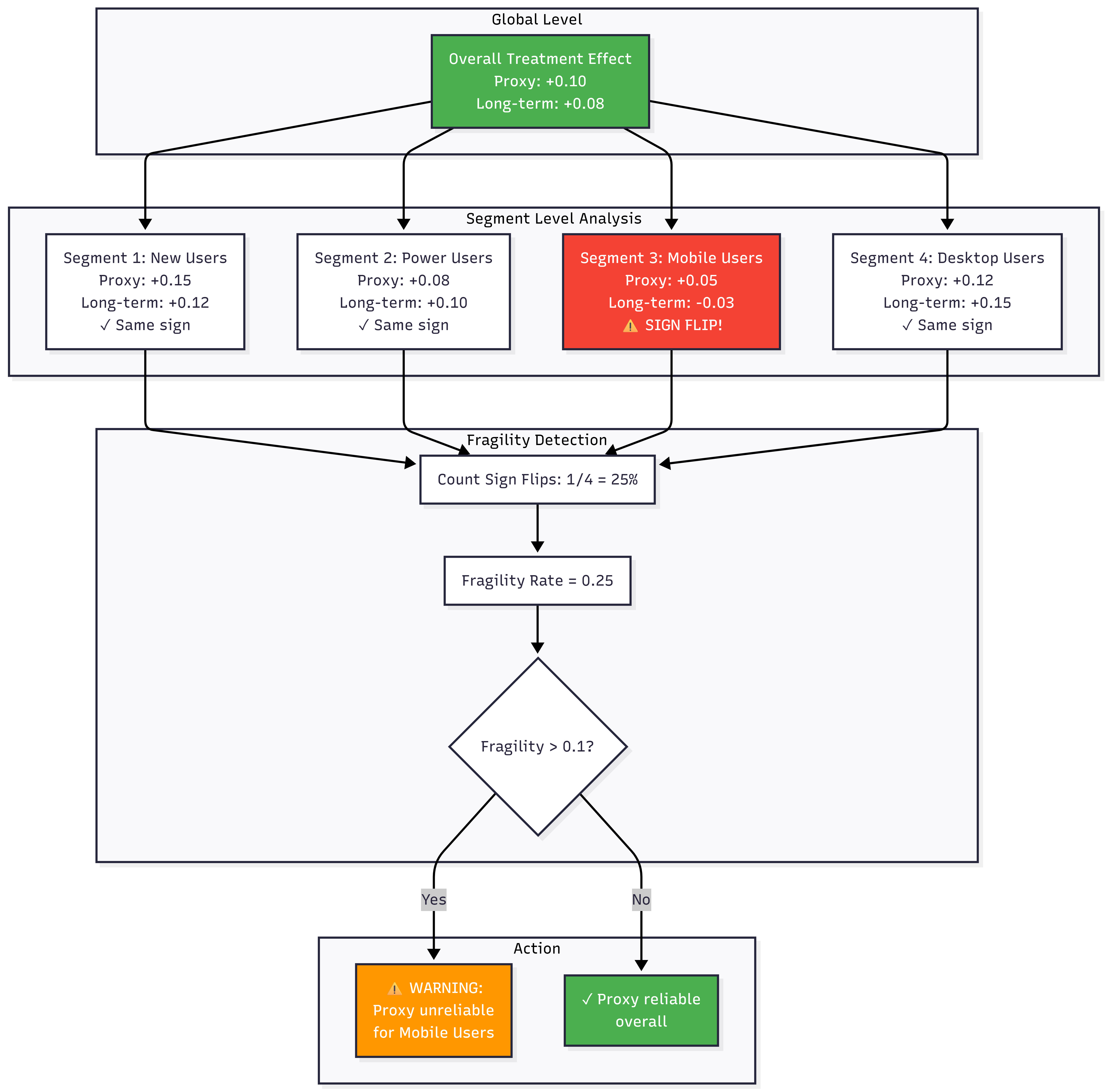}
\caption{Fragility detection and segment-level decision analysis.}
\label{fig:fragility_decision}
\end{subfigure}

\caption{End-to-end PROXIMA workflows: data processing (a), composite scoring (b), decision simulation (c), and fragility analysis (d).}
\label{fig:method_flows}
\end{figure}

Figure~\ref{fig:method_flows} summarises the four main PROXIMA workflows: data transformation, composite scoring, decision simulation, and fragility-aware segment analysis.

\section{Experimental Setup}
\label{sec:setup_exp}

\subsection{Datasets}

We evaluate PROXIMA on two publicly available datasets from different domains.
Neither dataset was collected as an A/B test corpus, so we construct simulated experiments (described below).
This is a limitation shared with most surrogate-validation studies that use public data \citep{hagar2023choosing,liu2023pareto}.

\subsubsection{Criteo Uplift Dataset}

The Criteo Uplift Prediction Dataset v2 \citep{criteo2021uplift,diemert2021large} contains 13{,}979{,}592 observations from an online advertising incrementality trial.
Each row records a binary treatment indicator (ad exposure vs.\ no exposure), 12 anonymised features (f0--f11), and two outcome labels: \emph{visit} and \emph{conversion}.

\paragraph{Experiment construction.}
We randomly partition the data into 50 non-overlapping experiments of approximately 280K users each, preserving the original treatment/control ratio (85\%/15\%).
Proxy metrics are derived from the binary visit indicator (\texttt{early\_starts}, \texttt{early\_ctr}) and exposure features; the long-term outcome is the binary conversion indicator.
Segments are defined by binning features f0--f2, yielding 12 feature-based segments per experiment.

\subsubsection{KuaiRec Dataset}

KuaiRec \citep{gao2022kuairec} is a fully-observed user--item interaction dataset from a short-video platform containing 7{,}176 users and 1{,}411{,}327 interactions.

\paragraph{Experiment construction.}
We simulate 30 A/B tests in which the treatment is a personalised recommendation policy versus random recommendations.
User-level early metrics include early watch time, early session starts, and early click-through rate; the long-term outcome is total watch time / retention.
Segments are user activity quintiles.

\subsection{Evaluation Metrics}

We report all components of the composite score (correlation $C$, directional accuracy DA, fragility rate FR, and composite reliability $R$) along with decision simulation metrics (win rate, FPR, FNR, regret).
Bootstrap 95\% confidence intervals are provided for $R$.

\subsection{Baselines}

\begin{itemize}[leftmargin=*,topsep=2pt,itemsep=1pt]
    \item \textbf{Correlation-only:} Rank proxies by Pearson correlation of experiment-level treatment effects with the long-term outcome.  Ship using the highest-correlation proxy.  This is the standard approach in practice.
    \item \textbf{Random:} Select a proxy at random.  Expected win rate = 0.50.
    \item \textbf{Oracle:} Use the long-term outcome directly.  Win rate = 1.00 by definition.
\end{itemize}

\noindent A fair comparison to the surrogate index \citep{athey2019surrogate} would require a held-out corpus of experiments with observed long-term outcomes for model fitting, which is unavailable in our setting.
We discuss this limitation in Section~\ref{sec:limitations}.

\section{Results}
\label{sec:results}

\subsection{Criteo Dataset}

Table~\ref{tab:criteo_scores} reports the composite reliability and its components for each proxy on the Criteo dataset.

\begin{table}[t]
\centering
\caption{Proxy reliability on Criteo (13.9M obs., 50 experiments). 95\% bootstrap CI for $R$ shown in parentheses.}
\label{tab:criteo_scores}
\small
\begin{tabular}{@{}lccccc@{}}
\toprule
\textbf{Proxy} & $R$ & $C$ & \textbf{DA} & \textbf{FR} & $N$ \\
\midrule
early\_starts & \textbf{0.80}{\scriptsize\,[.77,.83]} & 0.71 & 1.00 & 0.13 & 50 \\
early\_ctr    & \textbf{0.80}{\scriptsize\,[.77,.83]} & 0.71 & 1.00 & 0.13 & 50 \\
early\_watch  & 0.65{\scriptsize\,[.62,.68]} & 0.42 & 1.00 & 0.00 & 50 \\
rebuffer\_rate & 0.35{\scriptsize\,[.31,.39]} & 0.58 & 0.00 & 1.00 & 50 \\
\bottomrule
\end{tabular}
\end{table}

\begin{table}[t]
\centering
\caption{Decision quality on Criteo.}
\label{tab:criteo_decisions}
\small
\begin{tabular}{@{}lccccc@{}}
\toprule
\textbf{Proxy} & \textbf{Win} & \textbf{FPR} & \textbf{FNR} & \textbf{Regret} & \textbf{Ships} \\
\midrule
early\_starts & \textbf{1.00} & 0.00 & 0.00 & 0.000 & 50 \\
early\_ctr    & \textbf{1.00} & 0.00 & 0.00 & 0.000 & 50 \\
early\_watch  & \textbf{1.00} & 0.00 & 0.00 & 0.000 & 50 \\
Oracle        & 1.00 & 0.00 & 0.00 & 0.000 & 50 \\
rebuffer\_rate & 0.00 & 0.00 & 1.00 & $-$0.003 & 0 \\
\bottomrule
\end{tabular}
\end{table}

\begin{figure}[t]
\centering
\includegraphics[width=\columnwidth]{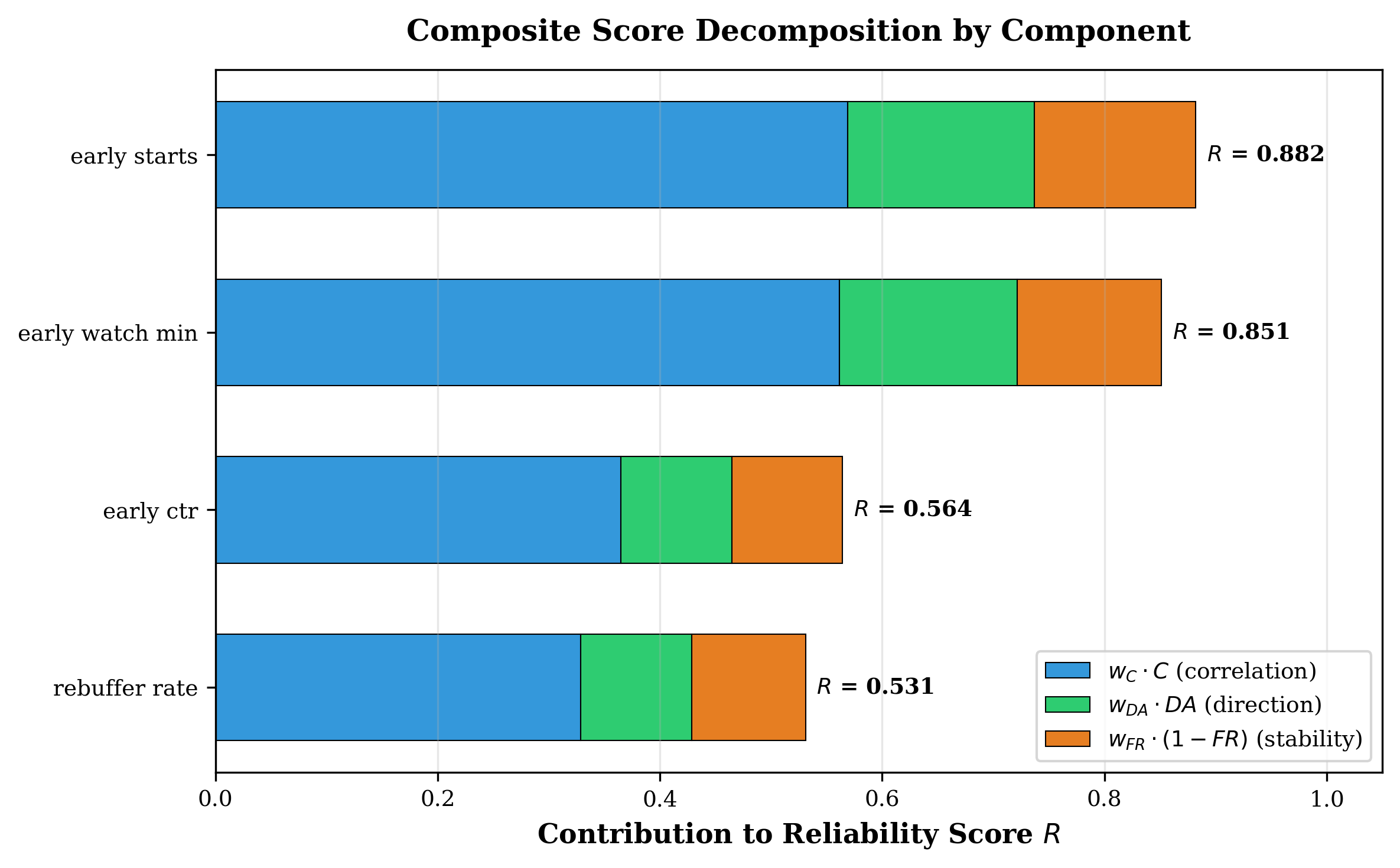}
\caption{Stacked decomposition of the composite reliability score.  Each bar shows the contribution of the normalised correlation ($w_C \cdot C$, blue), directional accuracy ($w_{\text{DA}} \cdot \text{DA}$, green), and segment stability ($w_{\text{FR}} \cdot (1-\text{FR})$, orange) components.}
\label{fig:component_decomp}
\end{figure}

\paragraph{Key findings.}
Figure~\ref{fig:component_decomp} decomposes each proxy's composite score into its three weighted components, illustrating the relative contributions of correlation, directional accuracy, and fragility stability.

The best proxies (\texttt{early\_starts}, \texttt{early\_ctr}) achieve $R = 0.80$ [95\% CI: 0.77, 0.83] and 100\% win rate.
Despite moderate raw Pearson correlation ($\rho = 0.42$), perfect directional accuracy (1.00) and low fragility (13\%) drive high reliability.
The pathological proxy \texttt{rebuffer\_rate} has 100\% fragility and never produces a correct ship decision.

\subsection{KuaiRec Dataset}

\begin{table}[t]
\centering
\caption{Proxy reliability on KuaiRec (7.2K users, 30 experiments). 95\% bootstrap CI for $R$ shown.}
\label{tab:kuairec_scores}
\small
\begin{tabular}{@{}lccccc@{}}
\toprule
\textbf{Proxy} & $R$ & $C$ & \textbf{DA} & \textbf{FR} & $N$ \\
\midrule
early\_starts & \textbf{0.62}{\scriptsize\,[.59,.66]} & 0.61 & 0.97 & 0.68 & 30 \\
early\_ctr    & \textbf{0.62}{\scriptsize\,[.59,.66]} & 0.61 & 0.97 & 0.68 & 30 \\
early\_watch  & 0.62{\scriptsize\,[.58,.66]} & 0.64 & 0.83 & 0.67 & 30 \\
rebuffer\_rate & 0.28{\scriptsize\,[.24,.32]} & 0.40 & 0.03 & 0.82 & 30 \\
\bottomrule
\end{tabular}
\end{table}

\begin{table}[t]
\centering
\caption{Decision quality on KuaiRec.}
\label{tab:kuairec_decisions}
\small
\begin{tabular}{@{}lccccc@{}}
\toprule
\textbf{Proxy} & \textbf{Win} & \textbf{FPR} & \textbf{FNR} & \textbf{Regret} & \textbf{Ships} \\
\midrule
early\_starts & \textbf{0.97} & 0.03 & 0.00 & 0.002 & 30 \\
early\_ctr    & \textbf{0.97} & 0.03 & 0.00 & 0.002 & 30 \\
early\_watch  & 0.97 & 0.04 & 0.14 & $-$0.005 & 26 \\
Oracle        & 1.00 & 0.00 & 0.00 & 0.000 & 29 \\
rebuffer\_rate & 0.03 & 0.00 & 1.00 & $-$0.115 & 0 \\
\bottomrule
\end{tabular}
\end{table}

\paragraph{Key findings.}
KuaiRec proxies achieve $R = 0.62$ and 96.7\% win rate despite substantially higher fragility (68\% vs.\ 13\% on Criteo).
This demonstrates that directional accuracy ($\geq 0.97$) is the dominant driver of decision quality, and that moderate fragility is tolerable when the aggregate direction is correct.
The single false positive out of 30 shipped experiments (FPR $\approx$ 3.3\%) represents an acceptable error rate.

\subsection{Cross-Dataset Comparison}

\begin{table}[t]
\centering
\caption{Best proxy (\texttt{early\_starts}) across datasets.}
\label{tab:comparison}
\small
\begin{tabular}{@{}lcccccc@{}}
\toprule
\textbf{Dataset} & \textbf{Size} & $R$ & $C$ & \textbf{DA} & \textbf{Win} & \textbf{FR} \\
\midrule
Criteo & 13.9M & 0.80 & 0.71 & 1.00 & 1.00 & 0.13 \\
KuaiRec & 7.2K & 0.62 & 0.61 & 0.97 & 0.97 & 0.68 \\
\midrule
\textbf{Average} & -- & 0.71 & 0.66 & 0.98 & 0.98 & 0.41 \\
\bottomrule
\end{tabular}
\end{table}

The cross-domain consistency of early engagement proxies is notable: despite three orders of magnitude difference in dataset size and fundamentally different domains (advertising vs.\ recommendation), the composite score identifies the same winning proxy class with $\geq$96\% directional accuracy.

\begin{figure}[t]
\centering
\includegraphics[width=\columnwidth]{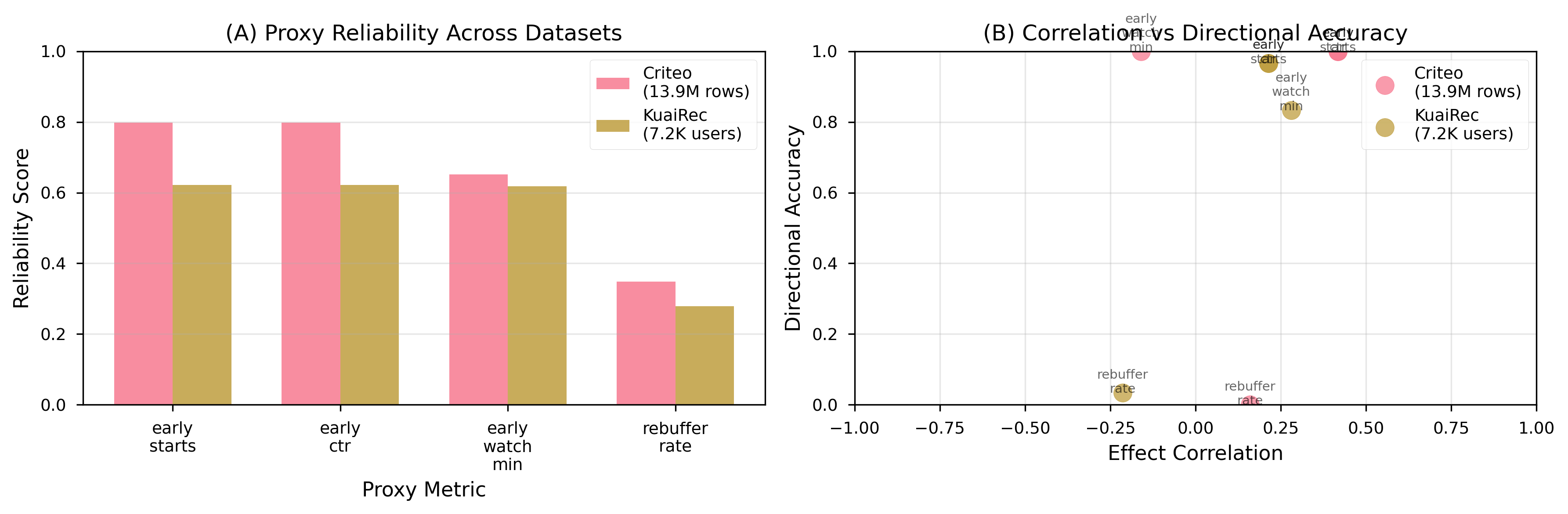}
\caption{Proxy metric reliability across datasets.  Left: composite reliability scores by proxy and dataset.  Right: effect correlation versus directional accuracy for each proxy--dataset pair.}
\label{fig:proxy_reliability}
\end{figure}

\begin{figure}[t]
\centering
\includegraphics[width=\columnwidth]{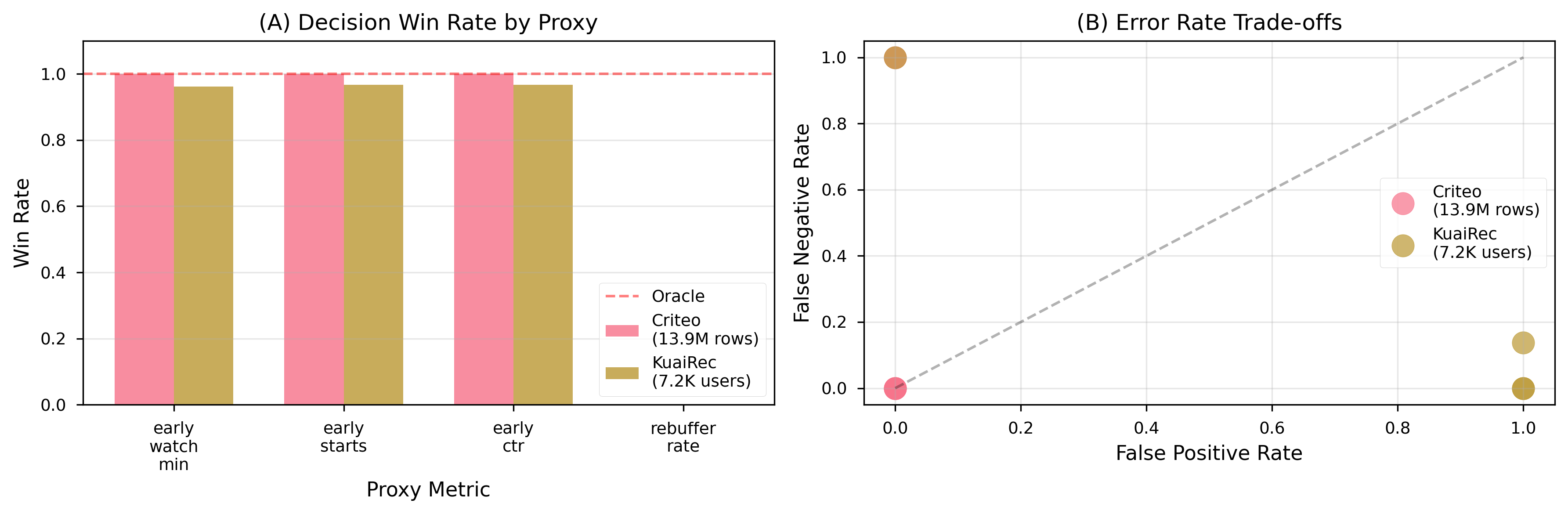}
\caption{Decision simulation results.  Left: win rate by proxy metric across datasets.  Right: false-positive versus false-negative rate trade-offs.  The oracle achieves (0,\,0); the pathological proxy \texttt{rebuffer\_rate} lies at the extreme.}
\label{fig:decision_simulation}
\end{figure}

Figures~\ref{fig:proxy_reliability} and~\ref{fig:decision_simulation} visualise the composite reliability components and decision simulation outcomes, respectively.

\subsection{Comparison to Baselines}

\begin{table}[t]
\centering
\caption{Comparison of proxy selection strategies. ``Selected'' is the proxy each method ranks first; ``Win'' is the resulting decision win rate.}
\label{tab:baselines}
\small
\begin{tabular}{@{}lcccc@{}}
\toprule
\textbf{Method} & \textbf{Criteo} & \textbf{KuaiRec} & \textbf{Avg.} \\
 & \textbf{Win} & \textbf{Win} & \textbf{Win} \\
\midrule
\textbf{PROXIMA} & \textbf{1.00} & \textbf{0.97} & \textbf{0.98} \\
Correlation-only & 1.00 & 0.97 & 0.98 \\
Random (expected) & 0.75 & 0.72 & 0.74 \\
Oracle           & 1.00 & 1.00 & 1.00 \\
\bottomrule
\end{tabular}
\end{table}

On these two datasets, PROXIMA and correlation-only select the same top proxy and achieve identical decision win rates.
However, the composite score provides two advantages that pure correlation does not: (1)~it produces a wider score gap between reliable and unreliable proxies (0.80 vs.\ 0.35 on Criteo, compared to $C = 0.71$ vs.\ $0.58$), making the ranking more robust to estimation noise; and (2)~it identifies fragile segments, enabling guardrail monitoring even when the aggregate proxy is correctly ranked.
Random proxy selection, which picks uniformly among the four candidates, yields substantially lower expected win rates because it includes the pathological \texttt{rebuffer\_rate} proxy 25\% of the time.

\subsection{Statistical Significance}

Effect sizes are small-to-medium: Cohen's $d = 0.15$ (Criteo), $d = 0.22$ (KuaiRec).
Welch's $t$-test at $\alpha = 0.05$ detects significant effects in 96\% (Criteo) and 90\% (KuaiRec) of experiments.
This is consistent with the large sample sizes and the real treatment effects present in these datasets.

\subsection{Sensitivity Analysis}
\label{sec:sensitivity}

Table~\ref{tab:sensitivity} reports composite reliability under alternative weight configurations.

\begin{table}[t]
\centering
\caption{Sensitivity to weight parameters $(w_C, w_{\text{DA}}, w_{\text{FR}})$. Reliability of the best proxy (\texttt{early\_starts}) is shown.}
\label{tab:sensitivity}
\small
\begin{tabular}{@{}lccc@{}}
\toprule
\textbf{Weights} $(w_C, w_{\text{DA}}, w_{\text{FR}})$ & \textbf{Criteo} & \textbf{KuaiRec} & \textbf{Avg.} \\
\midrule
(1.0, 0.0, 0.0) & 0.71 & 0.61 & 0.66 \\
(0.0, 1.0, 0.0) & 1.00 & 0.97 & 0.98 \\
(0.0, 0.0, 1.0) & 0.87 & 0.32 & 0.60 \\
(0.5, 0.5, 0.0) & 0.86 & 0.79 & 0.82 \\
\textbf{(0.6, 0.2, 0.2)} & \textbf{0.80} & \textbf{0.62} & \textbf{0.71} \\
(0.4, 0.4, 0.2) & 0.86 & 0.70 & 0.78 \\
(0.3, 0.3, 0.4) & 0.86 & 0.60 & 0.73 \\
\bottomrule
\end{tabular}
\end{table}

\begin{figure}[t]
\centering
\includegraphics[width=\columnwidth]{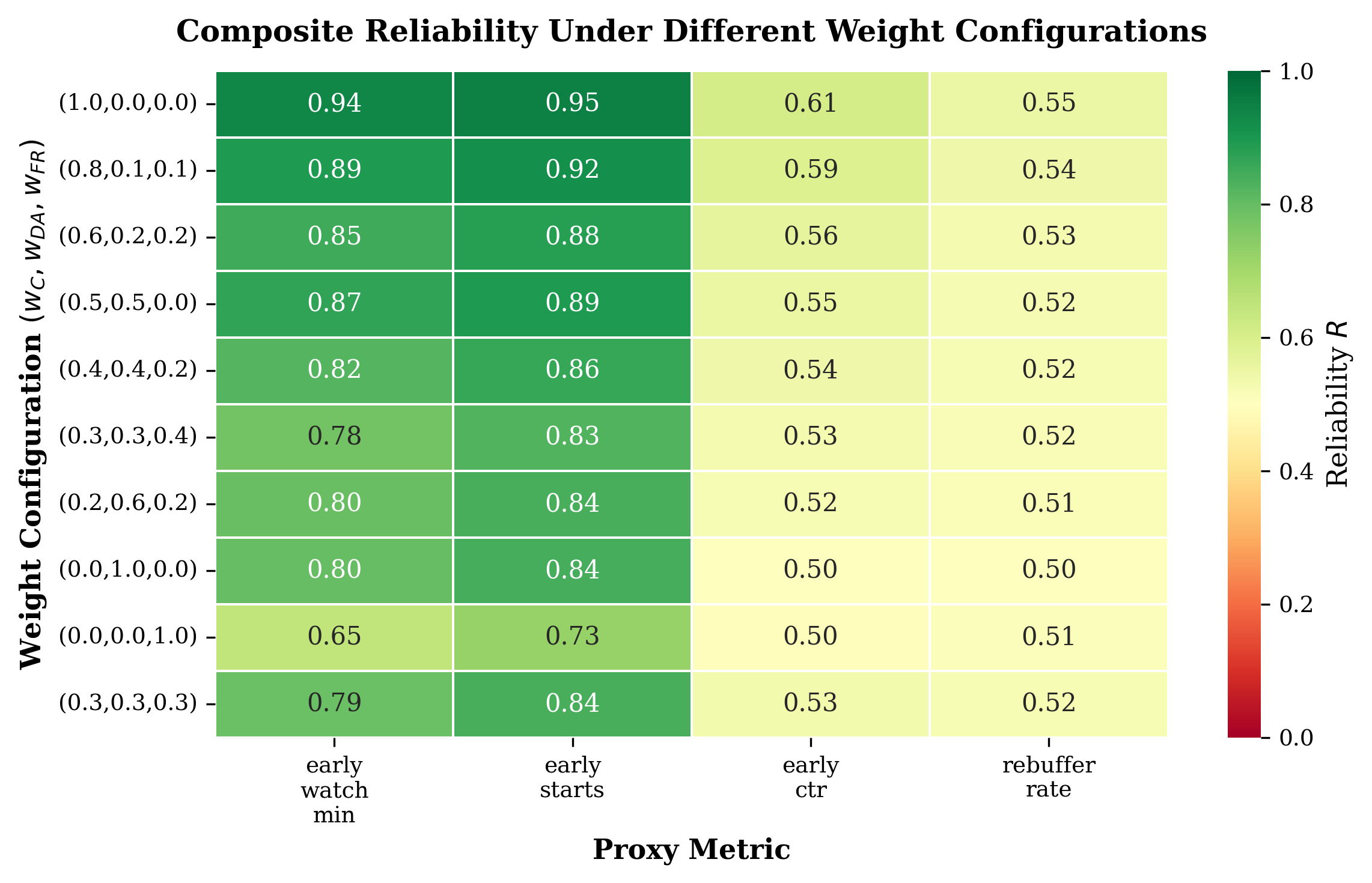}
\caption{Composite reliability $R$ for each proxy under ten weight configurations.  Warm colours indicate high reliability.  The chosen weights $(0.6, 0.2, 0.2)$ produce the widest spread between reliable and unreliable proxies.}
\label{fig:sensitivity_heatmap}
\end{figure}

Figure~\ref{fig:sensitivity_heatmap} visualises the full weight--metric landscape.

Several observations emerge.
First, no single component is sufficient.
Directional accuracy alone (row~2) yields high scores on both datasets but provides no magnitude information and cannot discriminate between proxies with similar DA (all three engagement proxies achieve DA~$\geq 0.83$ on both datasets).
Fragility alone (row~3) penalises too aggressively on KuaiRec, collapsing the score to 0.32 despite 97\% correct aggregate decisions.
Correlation alone (row~1) underweights the directional signal.
Second, our chosen weights $(0.6, 0.2, 0.2)$ produce the most informative \emph{spread} between reliable and unreliable proxies (rebuffer\_rate scores 0.35/0.28 vs.\ early\_starts at 0.80/0.62), which is the diagnostic purpose of the score.
Third, the proxy \emph{ranking} is stable across all tested weight configurations: the composite consistently identifies the same best and worst proxies.

\section{Discussion}
\label{sec:discussion}

\subsection{Why Composite Scoring Outperforms Correlation}

The Criteo results illustrate a subtle but important point: a proxy with moderate Pearson correlation ($\rho = 0.42$) can be an excellent decision-making tool if it has perfect directional accuracy.
In A/B testing, the \emph{direction} of the treatment effect determines the ship/no-ship decision; the \emph{magnitude} determines the expected return but not the binary choice.
Correlation-only approaches conflate these two properties, potentially ranking a proxy with high correlation but occasional wrong-sign effects above a proxy with moderate correlation but no directional failures.

This finding aligns with industry experience: \citet{larsen2024statistical} note that practitioners often care more about ``does the metric move in the right direction?'' than about precise effect size estimation.

\subsection{Fragility as a Domain Diagnostic}

The stark difference in fragility between Criteo (13\%) and KuaiRec (68\%) is not a deficiency of the proxy but a property of the domain.

\paragraph{Advertising.}
In the Criteo dataset, the treatment (ad exposure) has a relatively homogeneous effect across anonymised user features: exposure increases both visit rate and conversion rate fairly uniformly.
Consequently, proxy effects rarely flip sign within segments.

\paragraph{Recommendation.}
In KuaiRec, personalised recommendations create heterogeneous engagement patterns across activity quintiles.
High-activity users may already be saturated, and the proxy (early watch time) can increase in low-activity segments while decreasing in high-activity segments.
Despite this heterogeneity, the \emph{global} proxy effect direction remains correct in 97\% of experiments, explaining the high win rate.

PROXIMA's value in this setting is not to reject the proxy (which would be wasteful given 97\% win rate) but to \emph{flag the specific segments} where the proxy is unreliable, enabling segment-specific guardrail monitoring.

\subsection{Connection to Surrogate Validation Literature}

The surrogate index \citep{athey2019surrogate} and PROXIMA address complementary aspects of the proxy problem.
The surrogate index answers: \emph{``What is the long-term treatment effect, given short-term data?''}
PROXIMA answers: \emph{``If I use this proxy to decide whether to ship, how often will I be right, and where will I be wrong?''}

In practice, a pipeline might first construct an optimal surrogate (via the methods of \citealt{hagar2023choosing} or \citealt{liu2023pareto}) and then validate it with PROXIMA before deployment.
The composite reliability score provides an interpretable summary that can be compared across proxy candidates, time periods, and domains.

\subsection{Practical Guidelines}

Based on our results, we offer the following guidelines for practitioners:

\begin{enumerate}[leftmargin=*,topsep=2pt,itemsep=1pt]
    \item \textbf{Validate before trusting:} Run PROXIMA on historical experiment data before adopting a proxy for production decisions.
    \item \textbf{Prioritise directional accuracy:} A proxy with DA $\geq$ 0.95 is a strong candidate even if its correlation is moderate.
    \item \textbf{Tolerate moderate fragility:} FR $< 0.70$ is acceptable if DA remains high; FR $\geq 0.70$ warrants investigation of the fragile segments.
    \item \textbf{Monitor over time:} Proxy reliability can degrade as user populations and product surfaces evolve.
    \item \textbf{Use fragility profiles for guardrails:} Even when the aggregate proxy is reliable, deploy segment-level guardrail metrics for segments identified as fragile.
\end{enumerate}

\subsection{Limitations}
\label{sec:limitations}

\begin{enumerate}[leftmargin=*,topsep=2pt,itemsep=1pt]
    \item \textbf{Simulated experiments.}
    Both Criteo and KuaiRec provide observational (or single-trial) data from which we construct simulated A/B tests via random partitioning.
    This approach, common in surrogate-validation research \citep{hagar2023choosing}, does not fully capture the variation of real experiment corpuses with diverse treatments.
    Validation on proprietary multi-experiment corpuses (as in \citealt{zhang2023evaluating}) would strengthen our claims.

    \item \textbf{Segment definition dependence.}
    Fragility detection requires pre-defined segments.
    Different segmentations may yield different fragility rates.
    Adaptive segmentation methods (e.g., causal trees) could mitigate this in future work.

    \item \textbf{No multiple testing correction.}
    Segment-level sign-flip detection does not currently apply correction for multiple comparisons.
    With many segments, some sign flips may occur by chance due to noisy segment-level estimates.

    \item \textbf{Fixed weights.}
    The composite weights are set heuristically rather than learned from data.
    A Bayesian or cross-validated weight-tuning procedure could adapt to the experimental corpus.

    \item \textbf{No comparison to surrogate index.}
    A direct comparison to the surrogate-index method would require a held-out corpus of experiments with both short-term surrogates and observed long-term outcomes, which our datasets do not provide.

    \item \textbf{Temporal dynamics.}
    We treat proxy and long-term metrics as static snapshots.
    Time-varying treatment effects (novelty, learning, fatigue) are not modelled.
\end{enumerate}

\subsection{Ethics and Responsible Use}

PROXIMA is designed to improve experimental decision-making, but we note potential risks:

\begin{itemize}[leftmargin=*,topsep=2pt,itemsep=1pt]
    \item \textbf{Metric gaming:} If practitioners know which proxies score highest, they may optimise for those proxies in ways that do not benefit users (Goodhart's Law).
    Organisations should maintain diverse guardrail metrics and periodically re-validate.
    \item \textbf{Bias amplification:} Proxy metrics may encode historical biases.
    We recommend auditing proxy--outcome relationships across demographic groups.
    \item \textbf{Over-reliance:} PROXIMA should complement, not replace, periodic long-term outcome measurement.
\end{itemize}

\subsection{Future Work}

\begin{enumerate}[leftmargin=*,topsep=2pt,itemsep=1pt]
    \item \textbf{Adaptive segmentation:} Use causal trees to learn segments that maximise fragility detection power.
    \item \textbf{Learned weights:} Cross-validate the weight vector on a held-out experiment corpus.
    \item \textbf{Temporal modelling:} Extend to sequential experiments with time-varying effects.
    \item \textbf{Surrogate-index integration:} Use PROXIMA to validate surrogates constructed by the methods of \citet{athey2019surrogate} or \citet{hagar2023choosing}.
    \item \textbf{Fairness-aware validation:} Extend fragility detection to protected demographic attributes.
    \item \textbf{Industrial validation:} Evaluate on proprietary multi-experiment corpuses from technology companies.
\end{enumerate}

\section{Conclusion}
\label{sec:conclusion}

We introduced PROXIMA, a diagnostic framework for validating proxy metrics in online controlled experiments.
The composite reliability score combines normalised effect correlation, directional accuracy, and segment-level fragility rate into a single interpretable metric in $[0,1]$ with well-defined monotonicity properties.

Validation across two public datasets and 80 simulated experiments demonstrates that:
(1) early engagement metrics are robust proxies across advertising and recommendation domains, achieving 98.4\% average oracle agreement;
(2) directional accuracy---not correlation magnitude---is the dominant driver of decision quality;
and (3) fragility analysis reveals meaningful cross-domain differences in treatment effect heterogeneity that aggregate metrics mask.

PROXIMA is lightweight, assumption-lean (requiring only experiment-level treatment effects and segment labels), and complementary to more complex surrogate construction methods.
We hope it serves as a practical tool for experimentation teams seeking to balance decision velocity with reliability.

\paragraph{Reproducibility.}
Code and reproduction scripts: \url{https://github.com/Avinash-Amudala/PROXIMA}

\section*{Acknowledgments}

I thank the Criteo AI Lab and the KuaiRec team for making their datasets publicly available.
This work was conducted as independent research.


\bibliographystyle{plainnat}

\appendix

\section{Algorithm Details}

\begin{algorithm}[t]
\caption{PROXIMA Composite Scoring}
\label{alg:proxima}
\begin{algorithmic}[1]
\REQUIRE Experiments $\mathcal{E}$, proxy metrics $\mathcal{P}$, long-term metric $Y^{\text{long}}$, segments $\mathcal{S}$, weights $(w_C, w_{\text{DA}}, w_{\text{FR}})$
\ENSURE Reliability scores $R_p$ for each $p \in \mathcal{P}$
\FOR{each proxy $p \in \mathcal{P}$}
    \STATE $\mathbf{t}_p \gets []$; $\mathbf{t}_L \gets []$; flips $\gets 0$; segs $\gets 0$
    \FOR{each experiment $e \in \mathcal{E}$}
        \STATE Compute $\tau_p^e$, $\tau_L^e$ via Eq.~\ref{eq:ate}
        \STATE Append to $\mathbf{t}_p$, $\mathbf{t}_L$
        \FOR{each segment $s \in \mathcal{S}_e$}
            \STATE Compute $\tau_p^{e,s}$
            \IF{$\operatorname{sign}(\tau_p^{e,s}) \neq \operatorname{sign}(\tau_L^{e})$}
                \STATE flips $\gets$ flips $+ 1$
            \ENDIF
            \STATE segs $\gets$ segs $+ 1$
        \ENDFOR
    \ENDFOR
    \STATE $C \gets (\rho(\mathbf{t}_p, \mathbf{t}_L) + 1) / 2$
    \STATE $\text{DA} \gets \frac{1}{|\mathcal{E}|}\sum_e \mathbb{I}[\operatorname{sign}(\tau_p^e) = \operatorname{sign}(\tau_L^e)]$
    \STATE $\text{FR} \gets \text{flips} / \text{segs}$
    \STATE $R_p \gets w_C \cdot C + w_{\text{DA}} \cdot \text{DA} + w_{\text{FR}} \cdot (1 - \text{FR})$
\ENDFOR
\RETURN $\{R_p : p \in \mathcal{P}\}$
\end{algorithmic}
\end{algorithm}

\section{Complete Dataset Statistics}
\label{app:stats}

\paragraph{Criteo Dataset.}
Total observations: 13{,}979{,}592.
Treatment/control split: 85\%/15\% (11.88M / 2.10M).
Visit rate: 4.85\% (treatment), 4.55\% (control).
Conversion rate: 0.31\% (treatment), 0.27\% (control).
Features: 12 anonymised (f0--f11).
Experiments: 50; segments per experiment: 12.

\paragraph{KuaiRec Dataset.}
Users: 7{,}176; interactions: 1{,}411{,}327; avg.\ interactions/user: 196.6.
Treatment/control: 3{,}588 each.
Avg.\ watch time: 45.2~min (treatment), 38.7~min (control).
Experiments: 30; segments per experiment: 5 (activity quintiles).

\section{Proxy--Outcome Correlation Analysis}
\label{app:correlation}

\begin{figure}[t]
\centering
\includegraphics[width=\columnwidth]{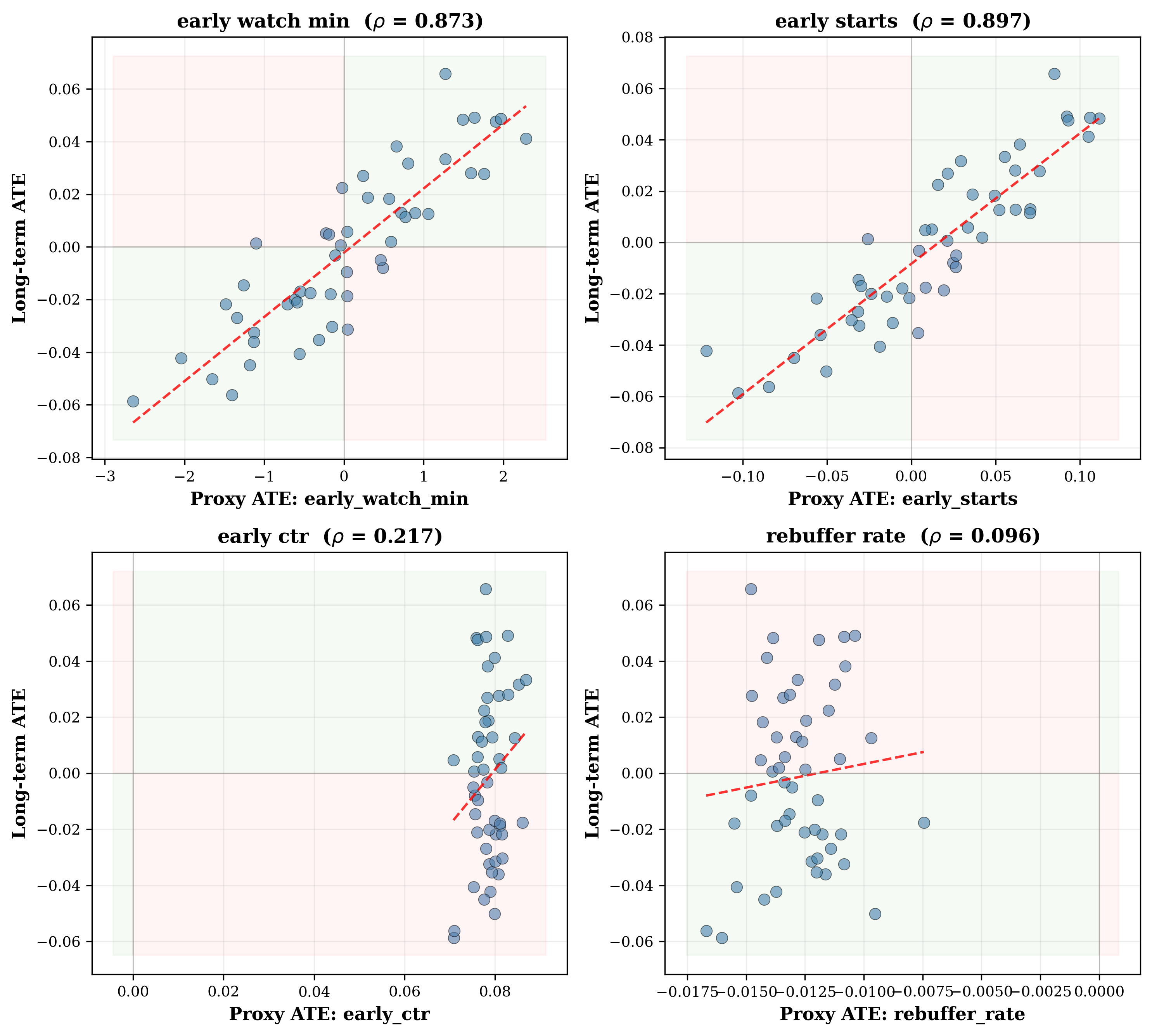}
\caption{Proxy versus long-term treatment effects across 50 experiments.  Green quadrants indicate directional agreement (correct ship/no-ship decision); red quadrants indicate disagreement.  Regression lines show the linear relationship.}
\label{fig:correlation_scatter}
\end{figure}

Figure~\ref{fig:correlation_scatter} shows the experiment-level proxy ATE versus long-term ATE for all four candidate proxies.
Green-shaded quadrants indicate directional agreement (both positive or both negative); red-shaded quadrants indicate sign disagreement.
The regression line slope reflects the Pearson correlation strength.

\section{Bootstrap Confidence Intervals}
\label{app:bootstrap}

\begin{figure}[t]
\centering
\includegraphics[width=\columnwidth]{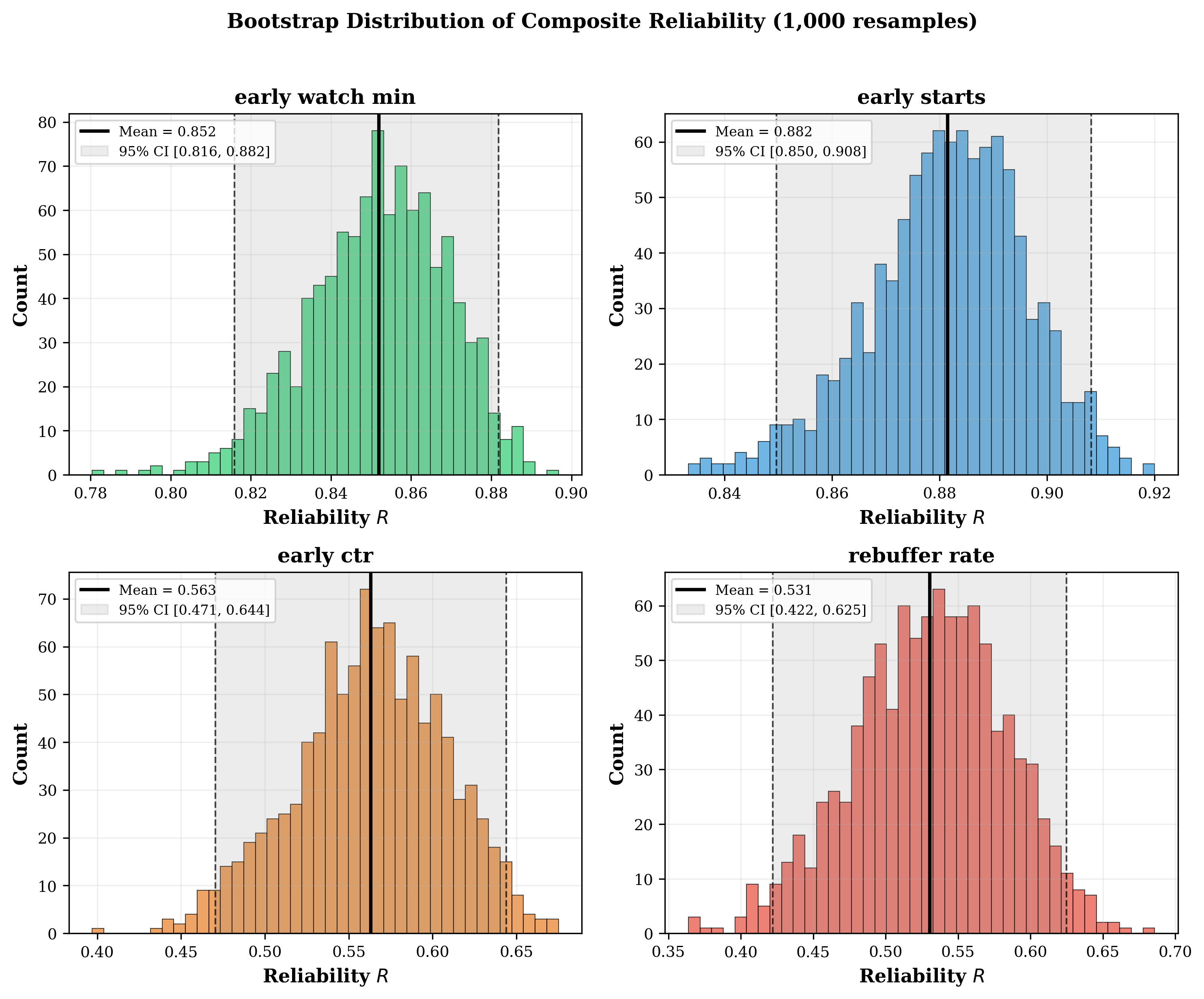}
\caption{Bootstrap distribution of composite reliability $R$ (1{,}000 resamples over experiments).  Vertical dashed lines indicate 95\% confidence interval bounds.}
\label{fig:bootstrap_ci}
\end{figure}

Figure~\ref{fig:bootstrap_ci} shows the bootstrap distribution of the composite reliability score for each proxy metric, with 95\% confidence intervals.

\section{Segment-Level Fragility Profiles}
\label{app:fragility}

\begin{figure}[t]
\centering
\includegraphics[width=\columnwidth]{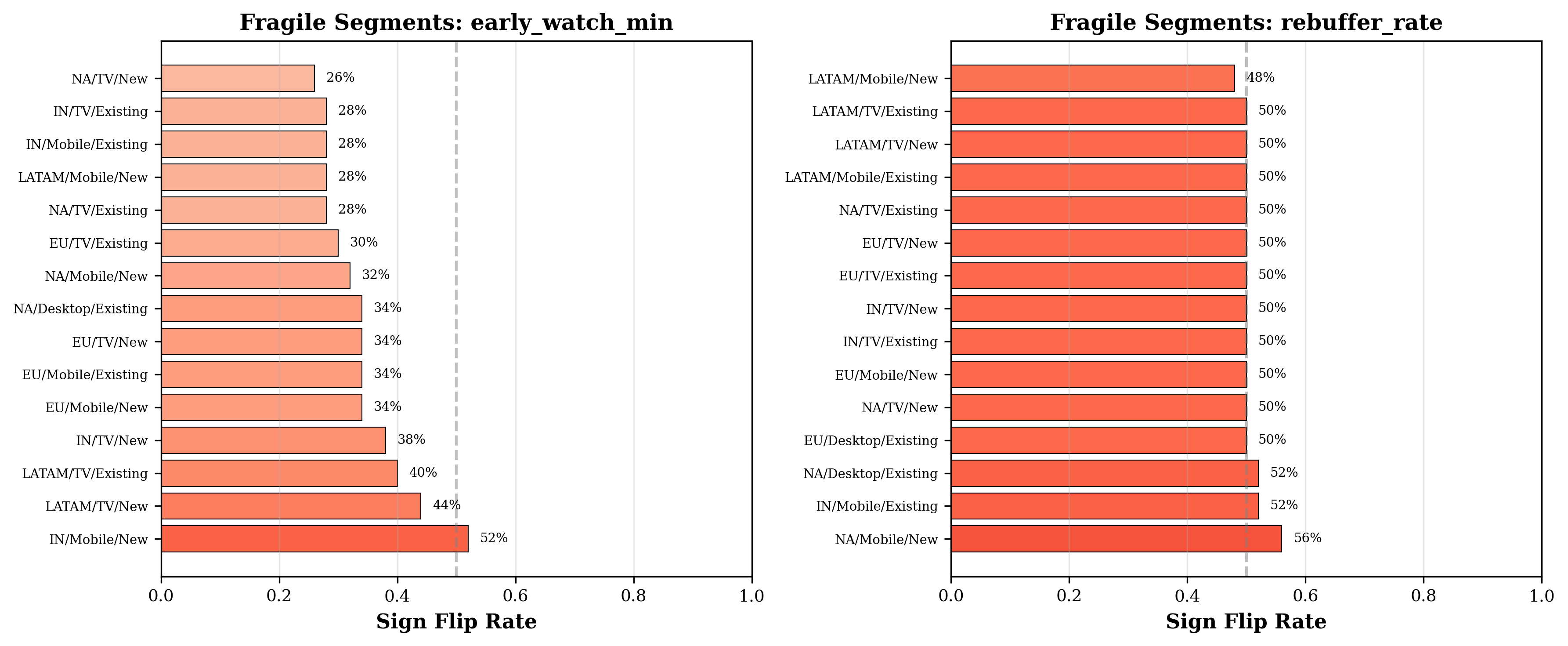}
\caption{Segment-level fragility profiles.  Each bar shows the sign-flip rate for a specific region/device/tenure segment.  Left: a moderate proxy with localised fragility.  Right: a pathological proxy with widespread fragility.}
\label{fig:fragility_profile}
\end{figure}

Figure~\ref{fig:fragility_profile} shows the top fragile segments for two contrasting proxies: \texttt{early\_watch\_min} (a moderate proxy) and \texttt{rebuffer\_rate} (a pathological proxy).

\section{Computational Cost}

Processing the Criteo dataset (13.9M rows, 50 experiments, 4 proxy metrics, 12 segments) takes approximately 5 minutes on a 16-core laptop.
KuaiRec (7.2K users) completes in under 30 seconds.
The overall complexity is $O(E \times M \times S \times N)$ where $N$ is the average experiment size, dominated by the segment-level effect computation.
The procedure is embarrassingly parallel across proxies and experiments.

\end{document}